\documentclass[preprint,aps,endfloats]{revtex4}
\usepackage[dvips]{graphicx}

\begin{document}

\title{Optimisation and Simulation of an Alternative nano-flash  Memory: the SASEM device.}

\author{C. Krzeminski$^{1}$, E. Dubois$^{1}$, X. Tang$^{2}$, N. Reckinger$^{2}$, A. Crahay$^{2}$ and V. Bayot$^{2}$}

\affiliation{$^{1}$ Institut Sup\'erieur d'Electronique de Micro\'electronique et de Nanotechnologies, UMR CNRS8520, D\'epartement ISEN, Avenue Poincar\'e, Cit\'e Scientifique, BP 69, 59652 Villeneuve d'Ascq, France.\\ $^{2}$ Microelectronics Laboratory, Universit\'e Catholique de Louvain, Place du Levant, 1348 Louvain-La-Neuve, Belgium.}
\email{christophe.krzeminski@isen.fr}

\begin{abstract}
Process simulation are performed in order to  simulate the full fabrication process of an alternative nano-flash memory in order to optimise it and to improve the understanding of the  dot storage formation. The influence of various parameters (oxidation temperature, nanowire shape) have been investigated.
\end{abstract}

\maketitle
\section*{INTRODUCTION.}
Since the conventional  floating gate device is believed to be  hardly scalable below the 65 nm technology node, alternative storage structures for nonvolatile memories are strongly needed. The feasibility of a silicon-on-insulator nano flash memory device based on the differential oxidation rate resulting from gradients in the arsenic doping concentration has previously been reported \cite{tang,tang2}. The key processes involved in the fabrication are arsenic implantation,  lithography and wet oxidation. The resulting device is a triangular MOSFET with a nanocrystal floating gate embedded in the gate oxide. Our objective is  to improve the reliability of the process and to ensure the presence of the memory dot for various conditions.  We investigate the wet oxidation step in details.  A clear understanding of the mechanisms of dot formation and of the influence of self-limited oxidation effects on the final device should be undertaken. The various geometrical parameters (e.g dot surface, dot-channel distance ...) of the nano-device  have been simulated as a function of the oxidation temperature and the duration of this oxidation step. 

\section*{FABRICATION OF THE SASEM DEVICE.}
We first brefly summarize  the  fabrication and the resulting \b{S}elf \b{A}ligned \b{S}ingle \b{E}lectron \b{M}emory (SASEM). A  SOI wafer with a 200 nm thick silicon overlay is used as a starting substrate. A high dose of  Arsenic is implanted (1 $\times$10$^{15}$cm$^{-2}$)  to create a very localized and a very doped area. Since the oxidation rate is doping dependent \cite{hoplum, hoplum2}, it will be possible  to realize an anisotropic oxidation. Next a thin oxide film and a nitride layer is deposited on the substrate. E-beam lithography and reactive ion etching is used to pattern the device.  Figure \ref{fig:mrs01} presents the layout of the device. The central part of the device consists of a  150 nm by 150 nm square. Two 100 nm wide constrictions  are defined in order to realise the connection between the source and drain regions and the central part. The various dimensions are defined such as a dot can be created in the central part of the device  and  is separated from the source/drain regions.  Finally, a wet oxidation step is performed in order to create the dot. Figure \ref{fig:mrs02} shows a SEM photography of the central part where a nano-floating gate is observed on top of the channel.

\section*{PROCESS SIMULATION ENVIRONMENT.}
The main steps of the SASEM process have been described and  the most critical step is the wet oxidation since it governs the creation of the silicon dot on top of the channel. Oxidation is  simulated  by a standard {\sc Deal} and {\sc Grove} generalised viscoelastic approach \cite{deal,Merz} using a commercial simulator \cite{DIOS}.  We use the {\sc Ho} and {\sc Plummer} model \cite{hoplum,hoplum2} in order to describe the enhancement of the oxide growth rate through the arsenic dopants (anisotropic oxidation). We used a  standard set of parameters for the oxidation model which is known to give reasonable results on various configurations \cite{Senez}. Parameters are not so far from the standard parameters of the simulation tool\cite{Senez}.  The objective is not a get a perfect agreement by adjusting the parameters on a specific experimental configuration of the SASEM device but to give raisonnable and interesting trends.   Moreover, we must keep in mind that this process is very complex to simulate  since the refinement of the grid must be preserved during the oxidation and the diffusion and segregation of arsenic have to be correctly described. We start with an anisotropic grid with a very high level of refinement in the region of high doping where the dot is expected to be created. This condition is very  important. If the  mesh describing the dot geometry is very relaxed, the study  of the evolution of the dot properties could  be  very problematic. To keep the quality of the initial grid during oxidation, very small timesteps are performed ( $\inf$ than 1ms) in order to generate  reduced and  smooth adaptation of the grid. 
The main problem of the  process simulation in general  is  the difficulty to  access to specific  geometrical information during the oxidation process. Furthermore, it is not fully adapted to the specificity of the SASEM memory cell. To overcome this problem,  a monitoring tool  has been developed. The flowchart of the different steps is presented in figure \ref{fig:export}. The monitoring tool  generates a complex script  with a large amount of commands. The basic idea is to perform a spatial sampling of the shape of the MESA-structure during the  oxidation step.  After the  process simulation, the monitoring tool is able to treat all the flow of data generated to  access to the evolution of the various parameters (dot length, dot width, dot-channel distance ...). We will see that  the coupling between the process simulator and the monitoring tool is very efficient and allows a broad range of simulations.

\section*{SIMULATIONS.}
\subsection*{Example of shape monitoring.}
Figure \ref{fig:monitor} presents an example of  the monitoring of the silicon shape during a wet oxidation at 800$^{\circ}$C. The silicon shape is represented for [0, 20, 40, 60, 80] minutes of wet oxidation. The profile for zero minute of wet oxidation is not strictly straight since a short dry oxidation step (2 minutes) is performed before the wet oxidation. As shown by the two figures, the oxidation is very fast in the region where the arsenic concentration is high (the maximum for the arsenic profile is located at 240 nm). The dot is created after 52 minutes in the configuration. As shown by the  figure  \ref{fig:monitor}, the dot is consumed mostly at the bottom where the arsenic concentration is large. The top of the dot is clearly well protected by the nitride mask. In comparison,  the bottom of the channel is clearly much oxidized.

\subsection*{Evolution of dimensional parameters.}
Various parameters have been studied. We report here only  the evolution of the dot properties  (figure \ref{fig:dot_length}).  The thickness at the dot creation is nearly the same $\sim$ 50nm. This result indicates that the position of the maximum of the arsenic concentration governs the initial size of the silicon dot. There is a sharp decrease at the time of dot creation and next  the dot size decreases almost linearly with a slope of about 0.5 nm/min. The evolution of the maximum dot width  is also reported  on the right figure.

\subsection*{Influence of the oxidation temperature on the final shape.}
The evolution of the dot shape as a function of the oxidation temperature has been simulated. Figure \ref{fig:shape1}  reports the evolution of  both the channel and the dot shape. The time of oxidation for each temperature  has been set just at dot creation. We clearly observe that the size of the dot is strongly reduced with an increasing temperature.  In figure {\ref{fig:shape1}, the maximal half dot width is about 30 nm at 750$^{\circ}$C and reduce to less than 10 nm for 950$^{\circ}$C. For higher temperature, no dot is created. Furthermore, the oxidation temperature  has clearly an impact on the shape that tends to be flat for a high temperature of oxidation. This result indicates that the oxidation temperature has an important influence on the final structure and that a wet low oxidation temperature  is recommended.

\section*{CONCLUSIONS}
Only a short part of the simulations performed has been described here. Thanks to an extensive set of simulations,  we have now access to the evolution  of many parameters and evolutions which otherwise would have required many experiments. More understanding of the dot formation is the key to improve the process. To conclude, we stress that the optimisation of the process is of crucial interest in order to  allow the downscaling of this alternative nano flash memory. The simulation and the definition of a strategy for scaling  the device  is the next step to adress.

\section*{ACKNOWLEDGMENTS.}
This research is supported by the European Union through the IST-2001-32674 SASEM project.

\newpage
\begin{figure}[tbp]   
\center
\begin{minipage}[]{6cm}
\hspace{1cm}
\includegraphics[width=6cm ,angle=-90]{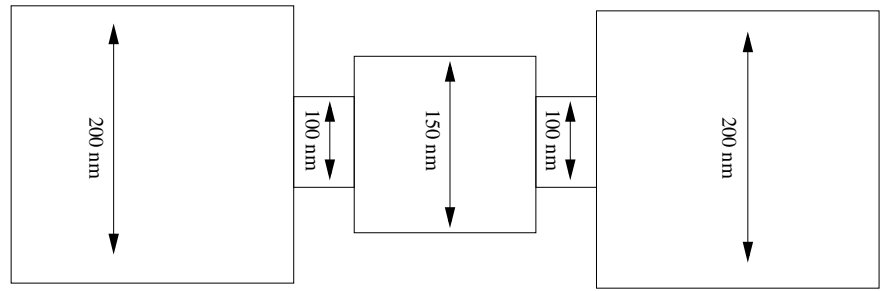}
\end{minipage}
\caption{Layout of the different regions of the SASEM device. }
\label{fig:mrs01}
\end{figure}
\newpage

\begin{figure}[tbp]   
\center
\begin{minipage}[c]{10cm}
\includegraphics[width=10cm,angle=0]{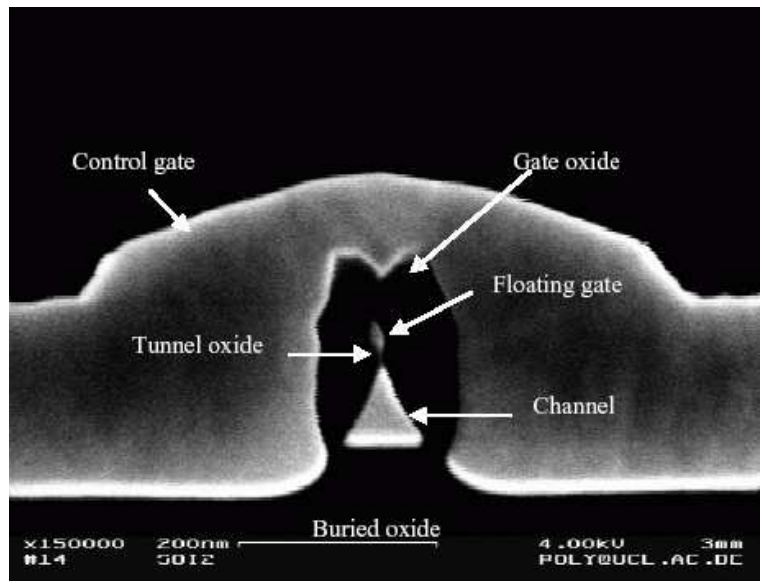}
\end{minipage}
\caption{SEM photography of the central part of the  nano-flash memory.}
\label{fig:mrs02}
\end{figure}

\newpage
\begin{figure}[]
\begin{minipage}[]{9cm}
\hspace{2cm}
\includegraphics[width=9cm ,angle=-90]{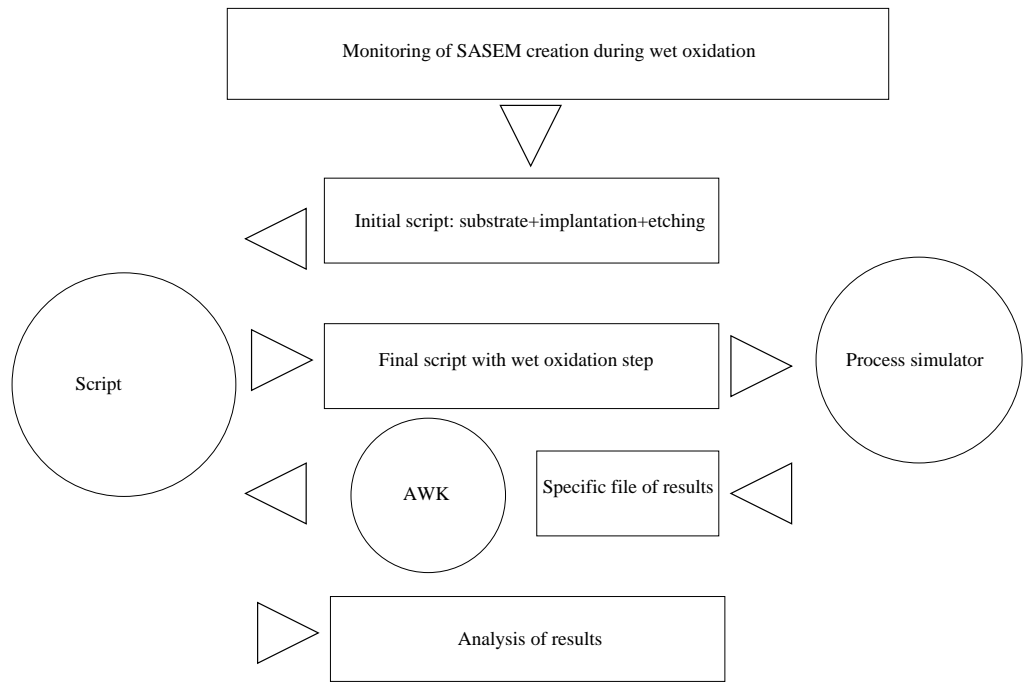}
\end{minipage}
\caption{Flowchart of the interaction between  the monitoring tool and the process simulator.}
\label{fig:export}
\end{figure}

\newpage
\begin{figure}[tbp]
\center
\begin{minipage}[c]{3cm}
\includegraphics[width=3cm]{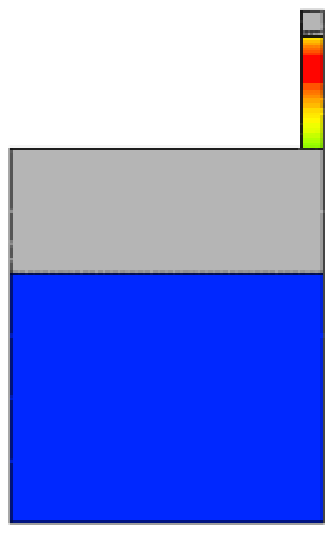}
\end{minipage}
\hspace{1.5cm}
\begin{minipage}[c]{7.5cm}
\includegraphics[width=7.5cm]{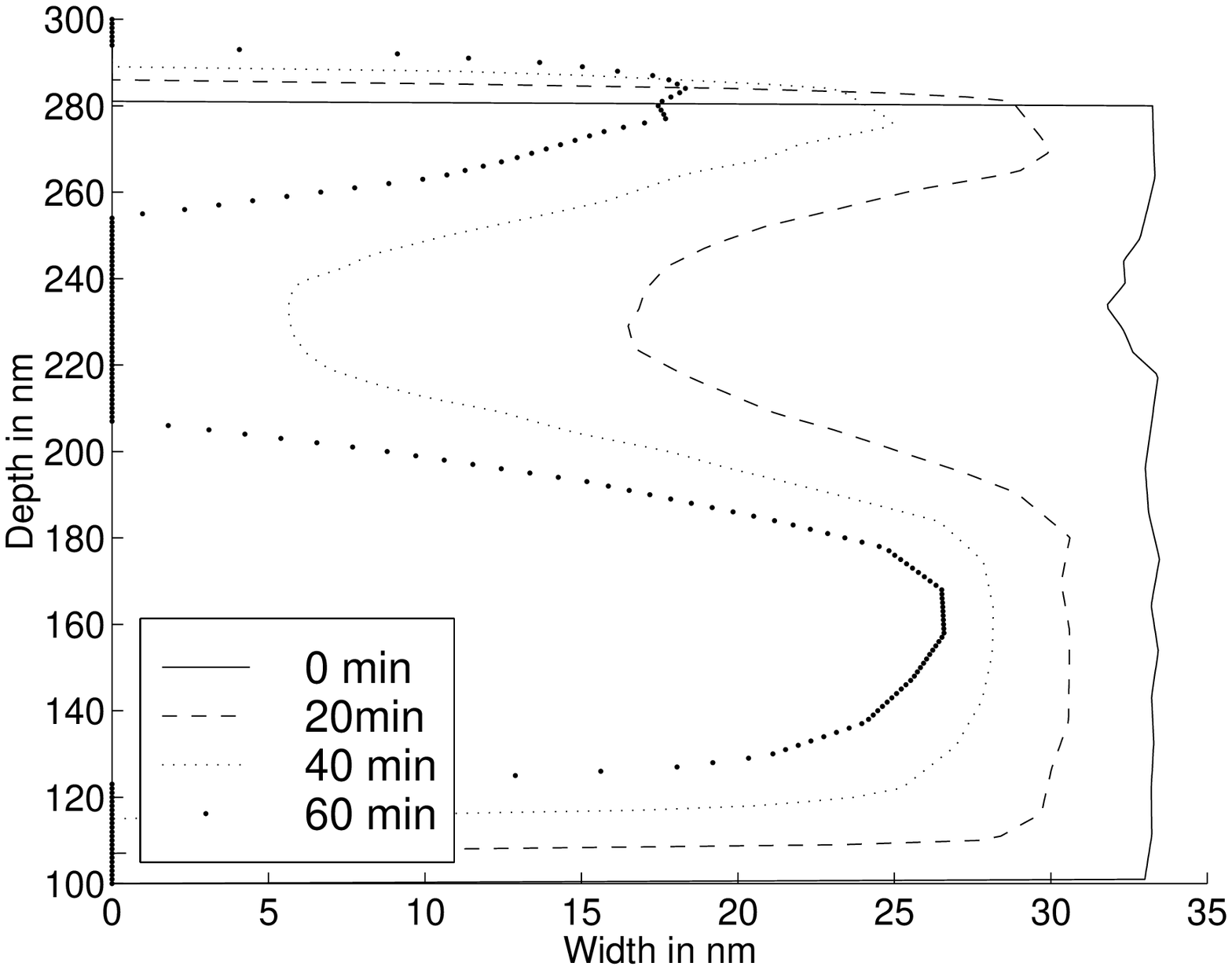}
\end{minipage}
\caption{ (Left) Initial configuration of the nanowire. (Right) Zoom on the evolution of the silicon  shape of the mesa structure during a wet oxidation step. The shape variations can be easily monitored thank to the monitoring tool.}
\label{fig:monitor}
\end{figure}

\newpage
\begin{figure}[tbp]
\center
\begin{minipage}[c]{7cm}
\includegraphics[width=7cm]{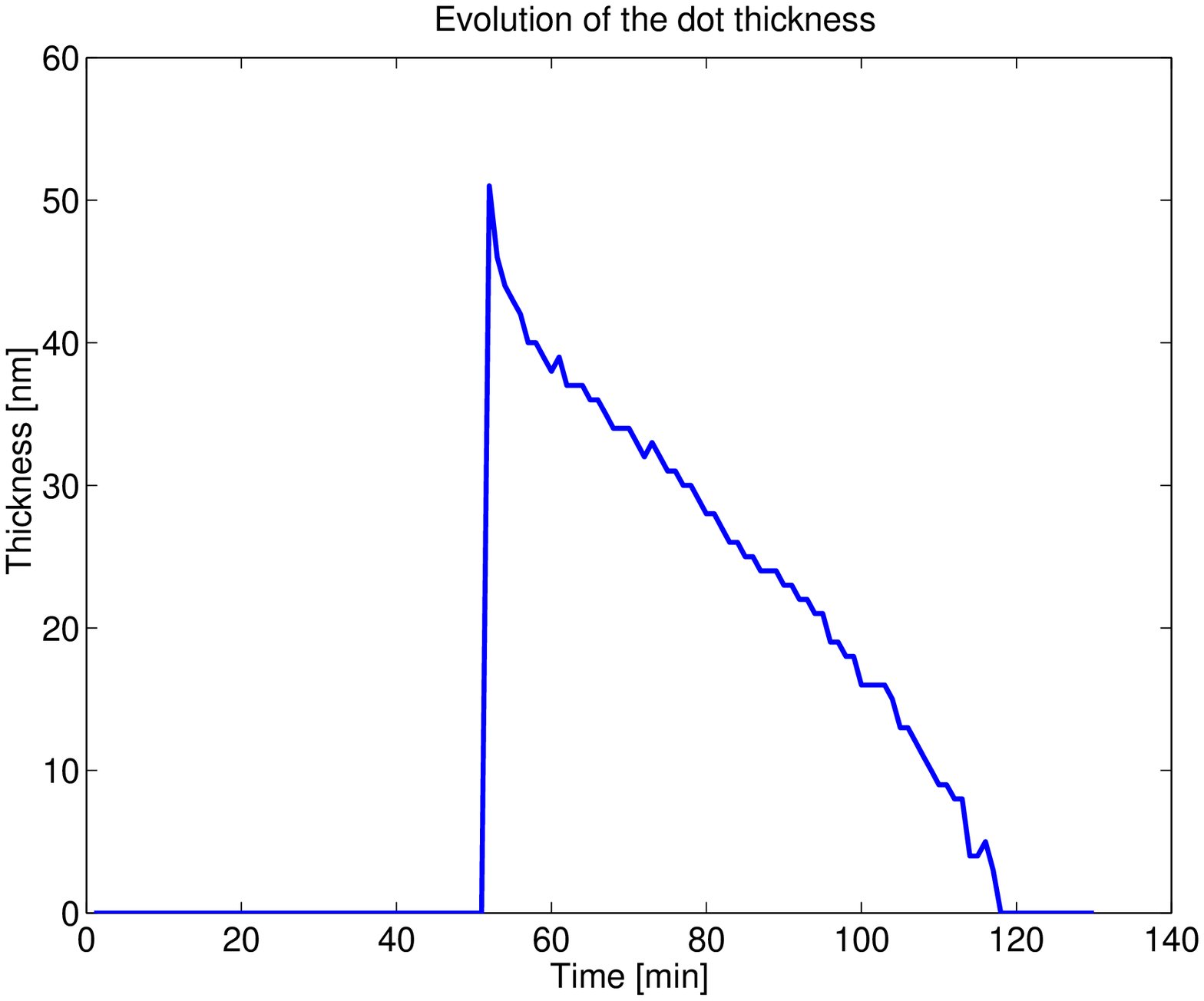}
\end{minipage}
\begin{minipage}[c]{7cm}
\includegraphics[width=7cm]{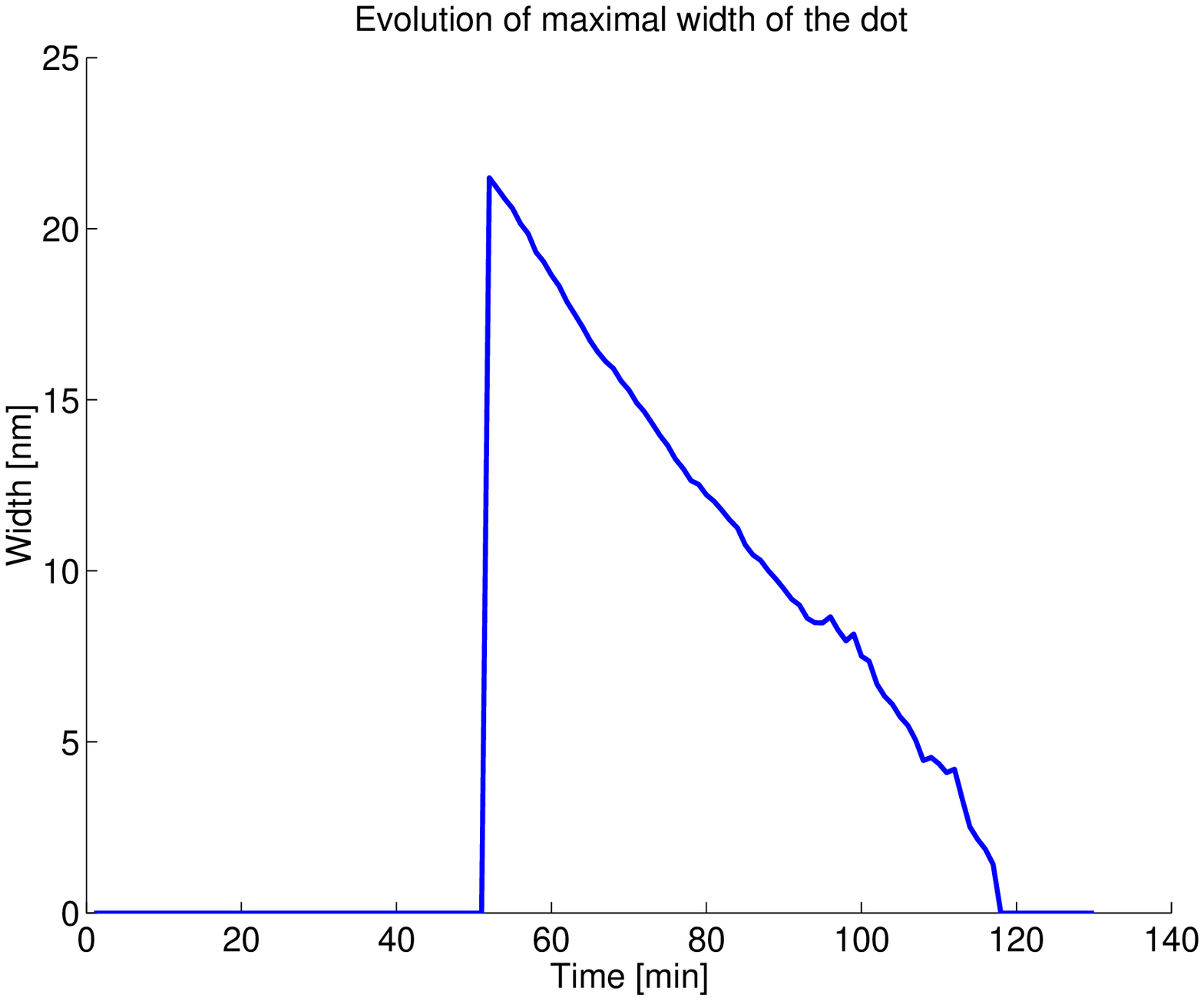}
\end{minipage}
\caption{(Left) Evolution of the thickness the dot  during the wet oxidation step. The thickness is close to 50 nm at the creation time and decreases almost linearly with the duration of the oxidation step.(Right) Evolution of the maximal width of the dot.}
\label{fig:dot_length}
\end{figure}

\begin{figure}[tbp]
\center
\begin{minipage}[c]{7cm}
\includegraphics[width=7cm]{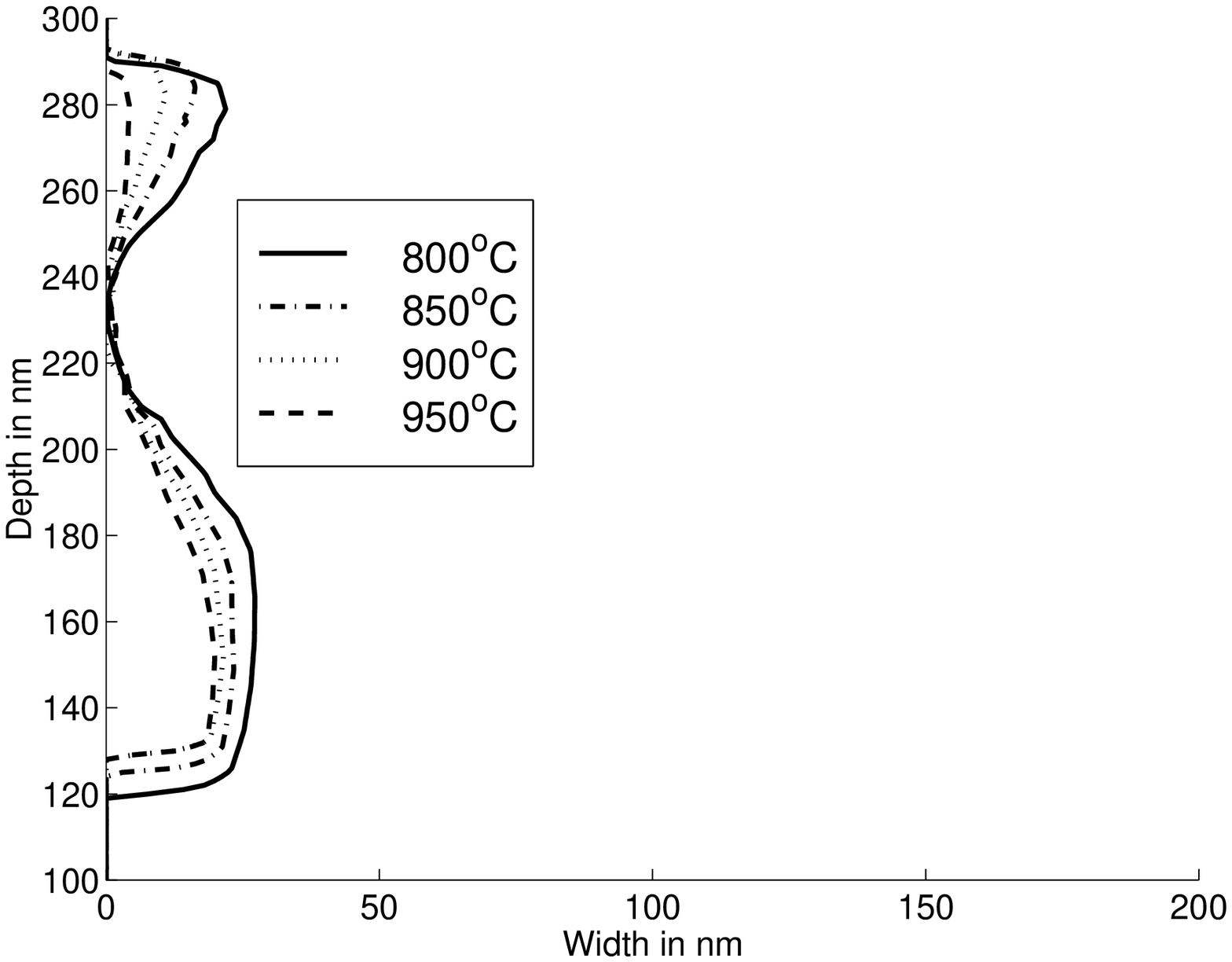}
\end{minipage}
\caption{Evolution of the MESA structure (only the silicon part)  and for various oxidation temperatures. The time of oxidation is set at  the dot creation time.}
\label{fig:shape1}
\end{figure}


\begin{thebibliography}{99}
\bibitem{tang}  X. {\sc Tang} et al., {\it Solid-State Electronics},{\bf 44}, p. 2259, 2000.
\bibitem{tang2} X. {\sc Tang},''Fabrication, characterization and simulation of SOI single-electron devices.'', {\it Phd. Thesis}, Universit\'e Catholique de Louvain-la-Neuve, April 2001.  
\bibitem{hoplum} C. P. {\sc Ho} and J. D. {\sc Plummer}, {\it J. Electrochem. Soc.: solid state science and technology}, {\bf Vol. 126}, No. 9,  p. 1516, 1979.
\bibitem{hoplum2} C. P. {\sc Ho} and J. D. {\sc Plummer}, {\it J. Electrochem. Soc.: solid state science and technology}, {\bf Vol. 126}, No. 9,  p. 1523, 1979.
\bibitem{deal} B. E. {\sc Deal} and A. S. {\sc Grove}, {\it Journal of Applied Physics}, {\bf Vol. 36}, No. 12, p. 3770, 1965.
\bibitem{Merz} W. {\sc Merz} and N. {\sc Strecker}, {\it Mathematical Methods in Applied Sciences},  p. 1165-1191, 1994.
\bibitem{DIOS} ISE DIOS, Process Simulator V6.0, ISE-TCAD.
\bibitem{Senez} V. {\sc Senez}, D. {\sc Collard}, B. {\sc Baccus} and J. {\sc Lebailly}, {\it  J. Appl. Phys.}, {\bf Vol. 43}, no. 5, p 720, 1996.


\end{thebibliography}
\end{document}